# Privacy Gain Based Multi-Iterative *k*-Anonymization to Protect Respondents' Privacy

Hitesh Chhinkaniwala[1*], Sanjay Garg[2]

[1]Assistant Professor, Computer Engineering Department, UV Patel College of Engineering,
Ganpat University, Mehsana, Gujarat, India

[2]Professor, Computer Engineering Department, Institute of Technology, Nirma University
Ahmedabad, Gujarat, India

## ABSTRACT

Huge volume of data from domain specific applications such as medical, financial, telephone, shopping records and individuals are regularly generated. Sharing of these data is proved to be beneficial for data mining application. Since data mining often involves data that contains personally identifiable information and therefore releasing such data may result in privacy breaches. On one hand such data is an important asset to business decision making by analyzing it. On the other hand data privacy concerns may prevent data owners from sharing information for data analysis. In order to share data while preserving privacy, data owner must come up with a solution which achieves the dual goal of privacy preservation as well as accuracy of data mining task mainly clustering and classification. Privacy Preserving Data Publishing (*PPDP*) is a study of eliminating privacy threats like linkage attack while preserving data utility by anonymizing data set before publishing. Proposed work is an extension to *k*-anonymization where Privacy Gain (*PrGain*) has been computed for selective anonymization for set of tuples. Classification and clustering characteristics of original data and anonymized data using proposed algorithm have been evaluated in terms of information loss, execution time, and privacy achieved. Algorithm has been processed against standard data sets and analysis shows that values for sensitive attributes are being preserved with minimal information loss.

## 1. INTRODUCTION

Databases today can range in size into the terabyte. Within these masses of data lies hidden information of strategic importance. The newest answer is data mining, which is being used both to increase revenues and to reduce costs. The potential returns are enormous. Innovative organizations worldwide are already using data mining to locate and appeal to higher-value customers, to reconfigure their product offerings to increase sales, and to minimize losses due to error or fraud. Data mining is a process that uses a variety of data analysis tools to discover patterns (finding interesting information) and relationships in data that may be used to make valid predictions.

With more and more information accessible in electronic forms and available on the web, and with increasingly powerful data mining tools being developed and put into use, there are increasing concerns that data mining may pose a threat to privacy and data security. However, it is important to note that most of the major data mining applications focus on the development of scalable algorithms and also do not involve personal data. The focus of data mining technology is on the discovery of general patterns, not on specific information regarding individuals. In this sense, we believe that the real privacy concerns are with unconstrained access of individual records, like credit card and banking applications, for example, which must access privacy-sensitive information. For those data mining applications that do involve personal data, in many cases, privacy required to be preserve. Data mining can be valuable in many applications, but due to no sufficient protection data may be abused for other goals. The main factor of privacy breaching in data mining is data misuse. In fact, if the data consists of critical and private characteristics and/or this technique is abused, data mining can be hazardous for individuals and organizations. Therefore, it is necessary to prevent revealing not only the personal confidential information but also the critical knowledge.

Privacy Preserving Data Mining (*PPDM*) has been emerged to address the privacy issues in data mining. Embedding privacy into data mining has been an active and an interesting research area. Several data mining techniques, incorporating privacy protection mechanisms, have been proposed based on different approaches. Recent research in the area of PPDM has devoted much effort to determine a trade-off between privacy and the need for knowledge discovery, which is crucial in order to improve decision-making processes and other human activities. PPDM helps to protect personal, proprietary or sensitive information, to enable collaboration between different data owners and also to comply with legislative policies.

## 2. RELATED WORK

There have been several methods developed by researchers in the database community that process records in a "group-based" manner, using information about specific local records globally to transform the records in a way which preserves specific privacy metrics. These modified records can then be published without fear of reconstruction by attacks. There is an assumption that certain fields of a record contain *quasi-identifiers* that uniquely identify an individual associated with the record, as well as *sensitive* attributes that must not be linked to the individual by an untrusted third party. Three variants of grouping-based methods (*k-anonymity*, *ℓ-diversity*, and *t-closeness*) have been proposed that rely on achieving the final state where $k$ records look exactly the same.

- *k-anonymity* and its variants

The *k-anonymity* model proposed by Samarati and Sweeney [1-3] achieves privacy by enforcing the constraint that every row of the released database should be indistinguishable from at least $k$ other rows with respect to a selected set of attributes called *quasi-identifiers*. This is usually achieved by suppressing (removing all or part of a field value), generalizing (using some pre-specified hierarchy of values), swapping values in the database. *k-anonymity* protects against identity disclosure, it does not provide sufficient protection against attribute disclosure. Homogeneity attack and back ground knowledge attack create challenges to *k-anonymity* approaches. A variant of *k-anonymity* known as *ℓ-diversity* was introduced by Machanavajjhala *et al.* [4]. It guarantees privacy in certain situations where *k-anonymity* does not, such as when there is little diversity in the sensitive attributes or when the adversary has some background information. Proposed method in [4] seems to be biased towards privacy at the cost of usability. The *t-closeness* model is a further enhancement on the concept *k-anonymity* and *ℓ-diversity*. One characteristic of the *ℓ-diversity* model is that it treats all values of a given attribute in a similar way irrespective of its distribution in the data. This is rarely the case for real data sets, since the attribute values may be much skewed. This may make it more difficult to create feasible *ℓ-diverse* representations. Often, an adversary may use background knowledge of the global distribution in order to make inferences about sensitive values in the data. Furthermore, not all values of an attribute are equally sensitive. For example, an attribute corresponding to a disease may be more sensitive when the value is positive, rather than when it is negative. *t-closeness* requires that the distribution of a sensitive attribute in any equivalence class is close to the distribution of the attribute in the overall data set [5].

- *Personalized privacy preservation*

A corporation may have very different constraints on the privacy of its records as compared to an individual, so we may wish to treat the records in a given data set very differently for anonymization purposes. This means that the value of $k$ for anonymization is not fixed but may vary with the record. A condensation based approach [6] has been proposed for PPDM in the presence of variable constraints on the privacy of the data records. This technique constructs groups of non-homogeneous size from the data, such that it is guaranteed that each record lies in a group whose size is at least equal to its anonymity level. Subsequently, pseudo-data are generated from each group so as to create a synthetic data set with the same aggregate distribution as the original data set. Another interesting model of personalized anonymity is discussed in [7] in which a person can specify the level of privacy for his or her sensitive values. This technique assumes that an individual can specify a node of the domain generalization hierarchy in order to decide the level of anonymity that he can work with.

- *Utility based privacy preservation*

The loss of information can also be considered a loss of utility for data mining purposes. The problem of utility based PPDM was first studied in [8]. The idea was to improve the curse of dimensionality by separately publishing marginal tables containing attributes which have utility, but are also problematic for privacy preservation purposes. The generalizations performed on the marginal tables and the original tables in fact do not need to be the same. A method for utility based data mining using local recoding was proposed in [9]. The approach is based on the fact that different attributes have different utility from an application point of view. Most anonymization methods are *global*, in which a particular tuple value is mapped to the same generalized value. In local recoding, the data space is partitioned into a number of regions, and the mapping of the tuple to generalize value is local to that region. Another direction for utility based PPDM is to anonymize the data in such a way that it remains useful for particular kinds of data mining or database applications. For example, in [10], a method has been proposed for *k-anonymization* using an information-loss metric as the utility measure. Such an approach is useful for the problem of classification.

### 3. PRIVACY GAIN BASED MULTI-ITERATIVE K-ANONYMIZATION

#### A. Framework

Fig. 1 describes basic framework for *PrGain* based *k*-anonymization. Each quasi-identifier has been generalized at minimum level to achieve privacy protection without compromising mush on data utility.

*Quasi-identifiers* providing maximum privacy have been selected and anonymized tuples have been marked and stored in data set $D'$. Remaining unanonymized tuples $D - D'$ follow same procedure with minimum subsequent level of *quasi-identifiers'* generalization. Process terminates when entire dataset $D$ is *k*-anonymized or there is no further anonymization possible from dimension table on $D - D'$ data set. To evaluate protection provided to respondents' privacy against information loss, well accepted Naïve Bayes classification algorithm has been used to test data set $D$ and *k*-anonymized data set $D'$.

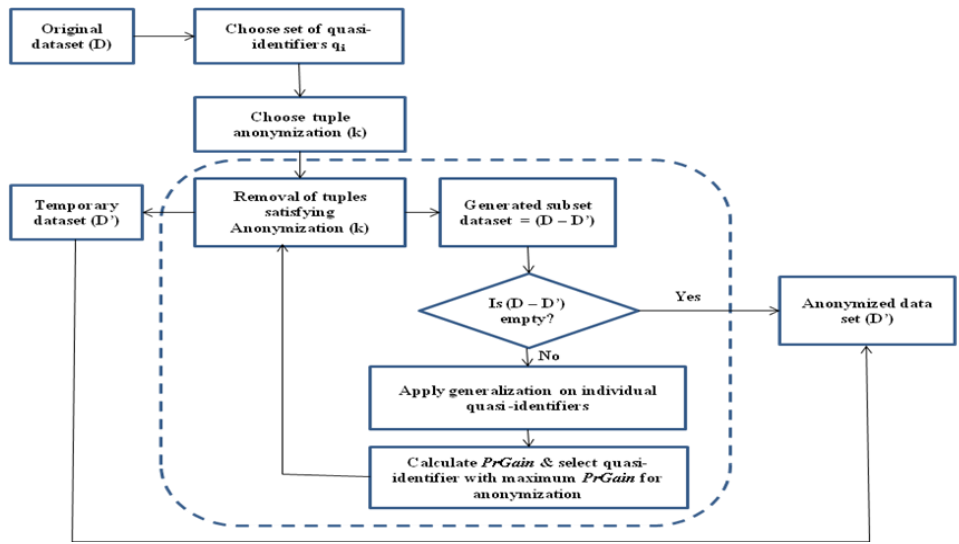

Fig. 1. **Framework for Privacy gain based multi-iterative *k*-anonymization**

#### B. Analyzing Framework

### Step 1: Calculate PrGain on original data set

| Non-Sensitive Data | | | Sensitive Data |
|---|---|---|---|
| **ZIP** | **Age** | **Gender** | **Condition** |
| 13053 | 28 | Male | Heart Disease |
| 13068 | 29 | Male | Heart Disease |
| 13068 | 21 | Female | Viral Infection |
| 13053 | 23 | Male | Viral Infection |
| 14853 | 50 | Female | Cancer |
| 14850 | 55 | Male | Heart Disease |
| 14850 | 47 | Female | Viral Infection |
| 14850 | 49 | Female | Viral Infection |
| 13053 | 31 | Male | Cancer |
| 13053 | 37 | Male | Cancer |
| 13068 | 36 | Female | Cancer |
| 13068 | 35 | Female | Viral Infection |
| 14850 | 32 | Male | Heart Disease |
| 13053 | 42 | Male | Hepatitis |
| 13068 | 40 | Female | Brochitis |
| 14850 | 28 | Male | Broken Arm |
| 13053 | 40 | Male | Viral Infection |
| 13053 | 25 | Female | Flu |
| 14853 | 51 | Female | Cancer |
| 13068 | 55 | Male | Heart Disease |

**Table 1 Original data set**

**Step 2: Generalization is applied on individual *quasi-identifier* $q_i \in Q$**

| Non-Sensitive Data | | | Sensitive Data |
|---|---|---|---|
| **ZIP** | **Age** | **gender** | **Condition** |
| 13053 | 28 | person | Heart Disease |
| 13068 | 29 | person | Heart Disease |
| 13068 | 21 | person | Viral Infection |
| 13053 | 23 | person | Viral Infection |
| 14853 | 50 | person | Cancer |
| 14850 | 55 | person | Heart Disease |
| 14850 | 47 | person | Viral Infection |
| 14850 | 49 | person | Viral Infection |
| 13053 | 31 | person | Cancer |
| 13053 | 37 | person | Cancer |
| 13068 | 36 | person | Cancer |
| 13068 | 35 | person | Viral Infection |
| 14850 | 32 | person | Cancer |
| 13053 | 42 | person | Hepatitis |
| 13068 | 40 | person | Brochitis |
| 14850 | 28 | person | Broken Arm |
| 13053 | 40 | person | Viral Infection |
| 13053 | 25 | person | Flu |
| 14853 | 51 | person | Cancer |
| 13068 | 55 | person | Heart Disease |

(a)

| Non-Sensitive Data | | | Sensitive Data |
|---|---|---|---|
| **ZIP** | **Age** | **Gender** | **Condition** |
| 13053 | 21-30 | Male | Heart Disease |
| 13068 | 21-30 | Male | Heart Disease |
| 13068 | 21-30 | Female | Viral Infection |
| 13053 | 21-30 | Male | Viral Infection |
| 14853 | 41-50 | Female | Cancer |
| 14853 | 51-60 | Male | Heart Disease |
| 14850 | 41-50 | Female | Viral Infection |
| 14850 | 41-50 | Female | Viral Infection |
| 13053 | 31-40 | Male | Cancer |
| 13053 | 31-40 | Male | Cancer |
| 13068 | 31-40 | Male | Cancer |
| 13068 | 31-40 | Female | Viral Infection |
| 14850 | 31-40 | Male | Heart Disease |
| 13053 | 41-50 | Female | Hepatitis |
| 13068 | 31-40 | Female | Brochitis |
| 14850 | 21-30 | Male | Broken Arm |
| 13053 | 31-40 | Male | Viral Infection |
| 13053 | 21-30 | Female | Flu |
| 14853 | 51-60 | Male | Cancer |
| 13068 | 51-60 | Male | Heart Disease |

(b)

| **ZIP** | **Age** | **Gender** | **Condition** |
|---|---|---|---|
| 1305* | 28 | Male | Heart Disease |
| 1306* | 29 | Male | Heart Disease |
| 1306* | 21 | Female | Viral Infection |
| 1305* | 23 | Male | Viral Infection |
| 1485* | 50 | Female | Cancer |
| 1485* | 55 | Male | Heart Disease |
| 1485* | 47 | Female | Viral Infection |
| 1485* | 49 | Female | Viral Infection |
| 1305* | 31 | Male | Cancer |
| 1305* | 37 | Male | Cancer |
| 1306* | 36 | Female | Cancer |
| 1306* | 35 | Male | Viral Infection |
| 1485* | 32 | Male | Heart Disease |
| 1305* | 42 | Male | Hepatitis |
| 1485* | 40 | Female | Brochitis |
| 1485* | 28 | Male | Broken Arm |
| 1305* | 40 | Male | Viral Infection |
| 1305* | 25 | Female | Flu |
| 1485* | 51 | Male | Cancer |
| 1306* | 55 | Male | Heart Disease |

(c)

Table 2 (a) <Gender[1]> anonymization $PrGain$ = NIL (b) <Age[1]> anonymization with $PrGain$ = 15% (c) <Zipcode[1]> anonymization $PrGain$ = NIL on Table 1

**Step 3: Generalization is applied for each quasi-identifier on unanonymized tuples from step 2**

| # | Non-Sensitive Data | | | Sensitive Data |
|---|---|---|---|---|
| | **ZIP** | **Age** | **Gender** | **Condition** |
| 1 | 13053 | 21-30 | Male | Heart Disease |
| 2 | 13068 | 21-30 | Male | Heart Disease |
| 3 | 13068 | 21-30 | Female | Viral Infection |
| 4 | 13053 | 21-30 | Male | Viral Infection |
| 5 | 14853 | 41-50 | Female | Cancer |
| 6 | 14853 | 51-60 | Male | Heart Disease |
| 7 | 14850 | 41-50 | Female | Viral Infection |
| 8 | 14850 | 41-50 | Female | Viral Infection |
| 9 | 13053 | 31-40 | Male | Cancer |
| 10 | 13053 | 31-40 | Male | Cancer |
| 11 | 13068 | 31-40 | Male | Cancer |
| 12 | 13068 | 31-40 | Female | Viral Infection |
| 13 | 14850 | 31-40 | Male | Heart Disease |
| 14 | 13053 | 41-50 | Female | Hepatitis |
| 15 | 13068 | 31-40 | Female | Brochitis |
| 16 | 14850 | 21-30 | Male | Broken Arm |
| 17 | 13053 | 31-40 | Male | Viral Infection |
| 18 | 13053 | 21-30 | Female | Flu |
| 19 | 14853 | 51-60 | Female | Cancer |

(a)

| # | Non-Sensitive Data | | | Sensitive Data |
|---|---|---|---|---|
| | **ZIP** | **Age** | **Gender** | **Condition** |
| 1 | 13053 | 21-30 | Male | Heart Disease |
| 2 | 13068 | 21-30 | Male | Heart Disease |
| 3 | 13068 | 21-30 | Female | Viral Infection |
| 4 | 13053 | 21-30 | Male | Viral Infection |
| 5 | 14853 | 41-50 | Female | Cancer |
| 6 | 14853 | 51-60 | Male | Heart Disease |
| 7 | 14850 | 41-50 | Female | Viral Infection |
| 8 | 14850 | 41-50 | Female | Viral Infection |
| 9 | 13053 | 31-40 | Male | Cancer |
| 10 | 13053 | 31-40 | Male | Cancer |
| 11 | 13068 | 31-40 | Male | Cancer |
| 12 | 13068 | 31-40 | Female | Viral Infection |
| 13 | 14850 | 31-40 | Male | Heart Disease |
| 14 | 13053 | 41-50 | Female | Hepatitis |
| 15 | 13068 | 31-40 | Female | Brochitis |
| 16 | 14850 | 21-30 | Male | Broken Arm |
| 17 | 13053 | 31-40 | Male | Viral Infection |
| 18 | 13053 | 21-30 | Female | Flu |
| 19 | 14853 | 51-60 | Female | Cancer |

(b)

Table 3 (a) <Age[1],Gender[0]> anonymization with $PrGain$ = 15% (b) <Age[1], Zipcode[0]> anonymization with $PrGain$ = 15% on Table 2 (b)

**Step 4: Generalization is applied for each quasi-identifier on unanonymized tuples from step 3. In case of equal PrGain both sets are considered.**

| # | Non-Sensitive Data | | | Sensitive Data |
|---|---|---|---|---|
| | **ZIP** | **Age** | **Gender** | **Condition** |
| 1 | 13053 | 21-30 | Person | Heart Disease |
| 2 | 13068 | 21-30 | person | Heart Disease |
| 3 | 13068 | 21-30 | person | Viral Infection |
| 4 | 13053 | 21-30 | Person | Viral Infection |
| 5 | 14853 | 41-50 | person | Cancer |
| 6 | 14853 | 51-60 | person | Heart Disease |
| 7 | 14850 | 41-50 | person | Viral Infection |
| 8 | 14850 | 41-50 | person | Viral Infection |
| 9 | 13053 | 31-40 | Male | Cancer |
| 10 | 13053 | 31-40 | Male | Cancer |
| 11 | 13068 | 31-40 | person | Cancer |
| 12 | 13068 | 31-40 | person | Viral Infection |
| 13 | 14850 | 31-40 | person | Heart Disease |
| 14 | 13053 | 41-50 | person | Hepatitis |
| 15 | 13068 | 31-40 | person | Brochitis |
| 16 | 14850 | 21-30 | person | Broken Arm |
| 17 | 13053 | 31-40 | Male | Viral Infection |
| 18 | 13053 | 21-30 | Person | Flu |
| 19 | 14853 | 51-60 | person | Cancer |
| 20 | 13068 | 51-60 | person | Heart Disease |

(a)

| # | Non-Sensitive Data | | | Sensitive Data |
|---|---|---|---|---|
| | **ZIP** | **Age** | **Gender** | **Condition** |
| 1 | 1305* | 21-30 | Male | Heart Disease |
| 2 | 1306* | 21-30 | Male | Heart Disease |
| 3 | 1306* | 21-30 | Female | Viral Infection |
| 4 | 1305* | 21-30 | Male | Viral Infection |
| 5 | 1485* | 41-50 | Female | Cancer |
| 6 | 1485* | 51-60 | Male | Heart Disease |
| 7 | 1485* | 41-50 | Female | Viral Infection |
| 8 | 1485* | 41-50 | Female | Viral Infection |
| 9 | 13053 | 31-40 | Male | Cancer |
| 10 | 13053 | 31-40 | Male | Cancer |
| 11 | 1306* | 31-40 | Female | Cancer |
| 12 | 1485* | 31-40 | Male | Viral Infection |
| 13 | 1485* | 31-40 | Male | Heart Disease |
| 14 | 1485* | 41-50 | Female | Hepatitis |
| 15 | 1306* | 31-40 | Female | Brochitis |
| 16 | 1485* | 31-40 | Male | Broken Arm |
| 17 | 13053 | 31-40 | Male | Viral Infection |
| 18 | 1305* | 21-30 | Male | Flu |
| 19 | 1485* | 51-60 | Female | Cancer |
| 20 | 1306* | 51-60 | Male | Heart Disease |

(b)

Table 4 (a) <$Age^1$,$Gender^1$> anonymization with *Pr Gain* = 45% (b) <$Age^1$, $Zipcode^1$> anonymization with *PrGain* = 15% on Table 3 (a)

**Step 5: Generalization is applied for each quasi-identifier on unanonymized tuples from step 4.**

| # | Non-Sensitive Data | | | Sensitive Data |
|---|---|---|---|---|
| | ZIP | Age | Gender | Condition |
| 1 | 13053 | 21-30 | Person | Heart Disease |
| 2 | 13068 | young | person | Heart Disease |
| 3 | 13068 | young | person | Viral Infection |
| 4 | 13053 | 21-30 | Person | Viral Infection |
| 5 | 14853 | mid age | person | Cancer |
| 6 | 14853 | older | person | Heart Disease |
| 7 | 14850 | mid age | person | Viral Infection |
| 8 | 14850 | mid age | person | Viral Infection |
| 9 | 13053 | 31-40 | Male | Cancer |
| 10 | 13053 | 31-40 | Male | Cancer |
| 11 | 13068 | 31-40 | person | Cancer |
| 12 | 13068 | 31-40 | person | Viral Infection |
| 13 | 14850 | mid age | person | Heart Disease |
| 14 | 13053 | mid age | person | Hepatitis |
| 15 | 13068 | 31-40 | person | Brochitis |
| 16 | 14850 | young | person | Broken Arm |
| 17 | 13053 | 31-40 | Male | Viral Infection |
| 18 | 13053 | 21-30 | Person | Flu |
| 19 | 14853 | older | person | Cancer |
| 20 | 13068 | older | person | Heart Disease |

(a)

| # | Non-Sensitive Data | | | Sensitive Data |
|---|---|---|---|---|
| | ZIP | Age | Gender | Condition |
| 1 | 13053 | 21-30 | Person | Heart Disease |
| 2 | 1306* | 21-30 | person | Heart Disease |
| 3 | 1306* | 21-30 | person | Viral Infection |
| 4 | 13053 | 21-30 | Person | Viral Infection |
| 5 | 1485* | 41-50 | person | Cancer |
| 6 | 1485* | 51-60 | person | Heart Disease |
| 7 | 1485* | 41-50 | person | Viral Infection |
| 8 | 1485* | 41-50 | person | Viral Infection |
| 9 | 13053 | 31-40 | Male | Cancer |
| 10 | 13053 | 31-40 | Male | Cancer |
| 11 | 13068 | 31-40 | person | Cancer |
| 12 | 13068 | 31-40 | person | Viral Infection |
| 13 | 1485* | 31-40 | person | Heart Disease |
| 14 | 1485* | 41-50 | person | Hepatitis |
| 15 | 13068 | 31-40 | person | Brochitis |
| 16 | 1485* | 21-30 | person | Broken Arm |
| 17 | 13053 | 31-40 | Male | Viral Infection |
| 18 | 13053 | 21-30 | Person | Flu |
| 19 | 1485* | 51-60 | person | Cancer |
| 20 | 1306* | 51-60 | person | Heart Disease |

(b)

Table 5 (a) <$Age^2$,$Gender^1$> anonymization with *PrGain* = 60% (b) <$Age^1$,$Gender^1$,$Zipcode^1$> anonymization with *PrGain* = 65% on Table 4 (a)

**Step 6: Generalization is applied for each quasi-identifier on unanonymized tuples from step 5.**

| # | Non-Sensitive Data | | | Sensitive Data |
|---|---|---|---|---|
| | ZIP | Age | Gender | Condition |
| 1 | 13053 | 21-30 | Person | Heart Disease |
| 2 | 1306* | young | person | Heart Disease |
| 3 | 1306* | young | person | Viral Infection |
| 4 | 13053 | 21-30 | Person | Viral Infection |
| 5 | 1485* | 41-50 | person | Cancer |
| 6 | 1485* | older | person | Heart Disease |
| 7 | 1485* | 41-50 | person | Viral Infection |
| 8 | 1485* | 41-50 | person | Viral Infection |
| 9 | 13053 | 31-40 | Male | Cancer |
| 10 | 13053 | 31-40 | Male | Cancer |
| 11 | 13068 | 31-40 | person | Cancer |
| 12 | 13068 | 31-40 | person | Viral Infection |
| 13 | 1485* | mid age | person | Heart Disease |
| 14 | 1485* | 41-50 | person | Hepatitis |
| 15 | 13068 | 31-40 | person | Brochitis |
| 16 | 1485* | young | person | Broken Arm |
| 17 | 13053 | 31-40 | Male | Viral Infection |
| 18 | 13053 | 21-30 | Person | Flu |
| 19 | 1485* | older | person | Cancer |
| 20 | 1306* | older | person | Heart Disease |

(a)

| # | Non-Sensitive Data | | | Sensitive Data |
|---|---|---|---|---|
| | ZIP | Age | Gender | Condition |
| 1 | 13053 | 21-30 | Person | Heart Disease |
| 2 | 130** | 21-30 | person | Heart Disease |
| 3 | 130** | 21-30 | person | Viral Infection |
| 4 | 13053 | 21-30 | Person | Viral Infection |
| 5 | 1485* | 41-50 | person | Cancer |
| 6 | 148** | 51-60 | person | Heart Disease |
| 7 | 1485* | 41-50 | person | Viral Infection |
| 8 | 1485* | 41-50 | person | Viral Infection |
| 9 | 13053 | 31-40 | Male | Cancer |
| 10 | 13053 | 31-40 | Male | Cancer |
| 11 | 13068 | 31-40 | person | Cancer |
| 12 | 13068 | 31-40 | person | Viral Infection |
| 13 | 148** | 31-40 | person | Heart Disease |
| 14 | 1485* | 41-50 | person | Hepatitis |
| 15 | 13068 | 31-40 | person | Brochitis |
| 16 | 148** | 21-30 | person | Broken Arm |
| 17 | 13053 | 31-40 | Male | Viral Infection |
| 18 | 13053 | 21-30 | Person | Flu |
| 19 | 148** | 51-60 | person | Cancer |
| 20 | 130** | 51-60 | person | Heart Disease |

(b)

Table 6 (a) <$Age^2$,$Gender^1$,$Zipcode^1$> anonymization with *PrGain* = 65% (b) <$Age^1$,$Gender^1$,$Zipcode^2$> anonymization with *PrGain* = 65% on Table 5 (b)

**Step 7: Generalization is applied for each quasi-identifier on unanonymized tuples from step 6. In case of equal PrGain both sets are considered.**

| # | Non-Sensitive Data | | | Sensitive Data |
|---|---|---|---|---|
| | ZIP | Age | Gender | Condition |
| 1 | 13053 | 21-30 | Person | Heart Disease |
| 2 | 130** | 21-30 | person | Heart Disease |
| 3 | 130** | 21-30 | person | Viral Infection |
| 4 | 13053 | 21-30 | Person | Viral Infection |
| 5 | 1485* | 41-50 | person | Cancer |
| 6 | 148** | 51-60 | person | Heart Disease |
| 7 | 1485* | 41-50 | person | Viral Infection |
| 8 | 1485* | 41-50 | person | Viral Infection |
| 9 | 13053 | 31-40 | Male | Cancer |
| 10 | 13053 | 31-40 | Male | Cancer |
| 11 | 13068 | 31-40 | person | Cancer |
| 12 | 13068 | 31-40 | person | Viral Infection |
| 13 | 148** | 31-40 | person | Heart Disease |
| 14 | 1485* | 41-50 | person | Hepatitis |
| 15 | 13068 | 31-40 | person | Brochitis |
| 16 | 148** | 21-30 | person | Broken Arm |
| 17 | 13053 | 31-40 | Male | Viral Infection |
| 18 | 13053 | 21-30 | Person | Flu |
| 19 | 148** | 51-60 | person | Cancer |
| 20 | 130** | 51-60 | person | Heart Disease |

(a)

| # | Non-Sensitive Data | | | Sensitive Data |
|---|---|---|---|---|
| | ZIP | Age | Gender | Condition |
| 1 | 13053 | 21-30 | Person | Heart Disease |
| 2 | 13*** | 21-30 | person | Heart Disease |
| 3 | 13*** | 21-30 | person | Viral Infection |
| 4 | 13053 | 21-30 | Person | Viral Infection |
| 5 | 1485* | 41-50 | person | Cancer |
| 6 | 14*** | 51-60 | person | Heart Disease |
| 7 | 1485* | 41-50 | person | Viral Infection |
| 8 | 1485* | 41-50 | person | Viral Infection |
| 9 | 13053 | 31-40 | Male | Cancer |
| 10 | 13053 | 31-40 | Male | Cancer |
| 11 | 13068 | 31-40 | person | Cancer |
| 12 | 13068 | 31-40 | person | Viral Infection |
| 13 | 14*** | 31-40 | person | Heart Disease |
| 14 | 1485* | 41-50 | person | Hepatitis |
| 15 | 13068 | 31-40 | person | Brochitis |
| 16 | 14*** | 21-30 | person | Broken Arm |
| 17 | 13053 | 31-40 | Male | Viral Infection |
| 18 | 13053 | 21-30 | Person | Flu |
| 19 | 14*** | 51-60 | person | Cancer |

(b)

Table 7 (a) $<Age^2, Gender^1, Zipcode^2>$ anonymization with *PrGain* = 65% on Table 6 (a)

(b) $<Age^1, Gender^1, Zipcode^3>$ anonymization with *PrGain* = 70% on Table 6 (b)

**Step 8: Generalization / is applied for each quasi-identifier on unanonymized tuples from step 7. In case of equal PrGain both sets are considered.**

| # | Non-Sensitive Data | | | Sensitive Data |
|---|---|---|---|---|
| | ZIP | Age | Gender | Condition |
| 1 | 13053 | 21-30 | Person | Heart Disease |
| 2 | 1**** | young | person | Heart Disease |
| 3 | 1**** | young | person | Viral Infection |
| 4 | 13053 | 21-30 | Person | Viral Infection |
| 5 | 1485* | 41-50 | person | Cancer |
| 6 | 1**** | older | person | Heart Disease |
| 7 | 1485* | 41-50 | person | Viral Infection |
| 8 | 1485* | 41-50 | person | Viral Infection |
| 9 | 13053 | 31-40 | Male | Cancer |
| 10 | 13053 | 31-40 | Male | Cancer |
| 11 | 13068 | 31-40 | person | Cancer |
| 12 | 13068 | 31-40 | person | Viral Infection |
| 13 | 1**** | mid age | person | Heart Disease |
| 14 | 1485* | 41-50 | person | Hepatitis |
| 15 | 13068 | 31-40 | person | Brochitis |
| 16 | 1**** | young | person | Broken Arm |
| 17 | 13053 | 31-40 | Male | Viral Infection |
| 18 | 13053 | 21-30 | Person | Flu |
| 19 | 1**** | older | person | Cancer |
| 20 | 1**** | older | person | Heart Disease |

Table 8 $<Age^2, Gender^1, Zipcode^4>$ anonymization with *PrGain* = 95%

## C. Algorithm

In proposed *PrGain* based algorithm, generalization has been applied for each *quasi-identifier*. *PrGain* has been computed for individual quasi-identifiers upon generalization based on dimension table. Quasi-identifier with maximum *PrGain* has been selected and tuples generalized have been marked *k*- anonymized. Anonymized tuples then removed from original data set

*D* and stored into new data set *D'*. For rest of unanonymized tuples in D, same process has been applied until *k*-anonymization has been achieved for entire data set *D* or generalization has been applied to all quasi-identifiers to the maximum level as mentioned in dimension table and no further generalization possible on unanonymized tuples.

**Procedure**    Privacy Gain based Multi-Iterative - Anonymization

**Input**    Data set $D$ to be anonymized

Tuple anonymization $k$

Set of *quasi-identifiers* $q_i \in Q$

Dimension table for each *quasi-identifiers* $q_i$.

**Output**    Anonymized data set $D'$

**Procedure:**

Let Data set $D$ contains $T$ transactions

$T^u$ = Un-anonymized transactions

$T^a$ = Anonymized transactions

Initially $T^u = T$

If (size $(T) < k$)

    Anonymization is not possible

Return

if (size $(T^u < k)$

    For each transactions in $T^u$ do

    Suppose there are $T'$ already anonymized transactions

$$T^a = T'$$

$$PrGain = \frac{(T - (T^u - T^a))}{T}$$

While (size $(T^u < k)$

    For each $q_i$ do

        For each $T_{qi}$ do

            If ($T_{qi}$ does not satisfy $k$-anonymity)

                Apply generalization on each $T_{qi}$

$$T_q^a = T_{qi}$$

$$(PrGain)_q = \frac{(T - (T^u - T_q^a))}{T}$$

$$PrGain = Max((PrGain)_{qi})$$

        Store $q_i$

$$T^a = T^a + T_{qi}$$

$$T^u = T^u - T^a$$

Store $T^a$ to $D'$

## 4.    RESULTS & ANALYSIS

Proposed algorithm for data anonymization has been developed using Java. Classification results have been compared over Naïve Bayes algorithm available within WEKA tool. To evaluate performance of proposed data anonymization algorithm, two standard data sets available on [11].

| Data set | Instances processed | Parameters | Original data sets | Anonymized data sets (q=2) | | |
|---|---|---|---|---|---|---|
| | | | | $k = 2$ | $k = 3$ | $k = 4$ |
| Adult [11] | 48K | Correctly classified (%) | 100 | 97.06 | 97.19 | 97.12 |
| | | Time taken to build model (*sec*) | 0.73 | 0.35 | 0.19 | 0.19 |
| Bank marketing [11] | 45K | Correctly classified (%) | 100 | 96.33 | 96.32 | 96.27 |
| | | Time taken to build model (*sec*) | 0.39 | 0.27 | 0.25 | 0.25 |

Table 9 Classification results: Original data sets vs. *k*-anonymized data sets (with q = 2)

| Data set | Instances processed | Parameters | Original data sets | Anonymized data sets (q=3) | | |
|---|---|---|---|---|---|---|
| | | | | $k = 2$ | $k = 3$ | $k = 4$ |
| Adult [11] | 48K | Correctly classified (%) | 100 | 96.44 | 96.36 | 96.45 |
| | | Time taken to build model (*sec*) | 0.73 | 0.15 | 0.16 | 0.15 |

| Bank marketing [11] | 45K | Correctly classified (%) | 100 | 93.22 | 94.03 | 93.63 |
| | | Time taken to build model (sec) | 0.39 | 0.22 | 0.20 | 0.19 |

Table 10 Classification results: Original data sets vs. *k*-anonymized data sets (with q = 3)

It has been observed while applying proposed *PrGain* based algorithm that respondents' information has been protected with minimal data loss. Further this approach is based on multi-iteration on non *k*-anonymized set of tuples greatly reduces processing time. Earlier proposed works were rescanning entire data set for anonymization, we have made an attempt to provide solution with lesser execution time by proposing Privacy Gain (*PrGain*) concept.

## 5. CONCLUSION

Protecting respondents' privacy while data mining creates challenges to data mining community. Two ways respondents' privacy can be preserve, one by modifying sensitive information itself and second by keeping sensitive information available for data mining while eliminating identifier attributes and generalizing and/or suppressing *quasi-identifiers*. We have coined new term called Privacy Gain while proposing algorithm to generalizing and/or suppressing *quasi-identifiers'* value. Selective anonymization has been applied based on *PrGain* computed in subsequent iteration. Proposed algorithm has an advantage of rescanning of only subset of original data set which is not *k*-anonymized. We have tested proposed *PrGain* based algorithm against standard data set and found that respondent privacy has been protected with small amount of data loss which occurs due to anonymization. Classification results with original data sets versus classification results obtained using anonymized data sets over Naïve Bayes algorithm have been compared. Anonymization parameter has been taken as *k* = 2, 3, 4 while number of *quasi-identifiers* have been set as q = 2, 3. It has been observed that proposed approach significantly reduces execution time as total data set rescan is no more required and provides better privacy without compromising much on data utility.